\pdfoutput=1
\documentclass[conference]{IEEEtran}
\IEEEoverridecommandlockouts

\usepackage{cite}
\usepackage{amsmath,amssymb,amsfonts}
\usepackage{algorithmic}
\usepackage{graphicx}
\usepackage{textcomp}
\usepackage{xcolor}
\def\BibTeX{{\rm B\kern-.05em{\sc i\kern-.025em b}\kern-.08em
    T\kern-.1667em\lower.7ex\hbox{E}\kern-.125emX}}

\usepackage[hidelinks]{hyperref}
\hypersetup{colorlinks=true,linkcolor=blue,citecolor=blue}
\usepackage[noabbrev,capitalize]{cleveref}

\usepackage{booktabs}

\usepackage{multirow}

\begin{document}

\title{Post-Training Quantization for Vision Mamba with k-Scaled Quantization and Reparameterization
\thanks{This work is supported in part by MediaTek, Inc., Hsinchu, Taiwan, under Grant MTKC-2023-1050, and in part by the Ministry of Science and Technology of Taiwan under Grant NSTC 113-2221-E-002-120-MY3 (Corresponding author: An-Yeu (Andy) Wu)}
}

% \author{\IEEEauthorblockN{Anonymous Authors}}
\author{
Bo-Yun Shi, Yi-Cheng Lo, An-Yeu (Andy) Wu and Yi-Min Tsai \\
\textit{Graduate Institute of Electronics Engineering, National Taiwan University, Taipei, Taiwan} \\ 
\{billy,alan\}@access.ee.ntu.edu.tw, andywu@ntu.edu.tw, yi-min.tsai@mediatek.com
}

\maketitle

\begin{abstract}
    The Mamba model, utilizing a structured state-space model (SSM), offers linear time complexity and demonstrates significant potential. Vision Mamba (ViM) extends this framework to vision tasks by incorporating a bidirectional SSM and patch embedding, surpassing Transformer-based models in performance. While model quantization is essential for efficient computing, existing works have focused solely on the original Mamba model and have not been applied to ViM. Additionally, they neglect quantizing the SSM layer, which is central to Mamba and can lead to substantial error propagation by naive quantization due to its inherent structure. In this paper, we focus on the post-training quantization (PTQ) of ViM. We address the issues with three core techniques: 1) a k-scaled token-wise quantization method for linear and convolutional layers, 2) a reparameterization technique to simplify hidden state quantization, and 3) a factor-determining method that reduces computational overhead by integrating operations. Through these methods, the error caused by PTQ can be mitigated. Experimental results on ImageNet-1k demonstrate only a 0.8–1.2\% accuracy degradation due to PTQ, highlighting the effectiveness of our approach.
\end{abstract}


\section{Introduction} \label{Sec::Intro}
The Mamba model \cite{gu2023mamba}\cite{dao2024transformers} is a recently proposed model that incorporates time-varying parameters into structured state-space model (SSM) \cite{gu2021efficiently} and introduces hardware-aware algorithms to achieve highly efficient training and inference. Due to its linear time complexity and promising performance, Mamba has garnered significant attention as a potential alternative to Transformers. Building on Mamba's success in NLP tasks, researchers have begun exploring its application in vision tasks. Among these efforts, Vision Mamba (ViM) \cite{2024_ViM} was the first to propose a bidirectional SSM, enabling Mamba, which is traditionally designed for sequential inputs, to gain spatial awareness for image processing. By integrating techniques such as patch embedding commonly used in Vision Transformers \cite{dosovitskiy2020image}, ViM successfully established a vision model with Mamba as its backbone, achieving performance that surpasses Transformer-based models in experiments.

However, similar to large models like CNNs and Transformers, increasing model sizes lead to significant computational demands, energy consumption, and memory requirements. To address this issue, quantization is one of the most common and effective techniques that can reduce models' size and computing cost. Post-Training Quantization (PTQ) is a technique that quantizes a pretrained model directly without retraining. It works by collecting the output range of the model when it processes a small amount of calibration data, and determine the quantization scale based on this information. The lower overhead and minimal data requirements make PTQ a favored quantization approach in practice. As a result, research in PTQ focuses on finding methods to reduce the accuracy drop associated with quantization.

While research on PTQ for Transformers is relatively mature, it remains scarce for Mamba. Prior works on Mamba's quantization include Mamba-PTQ \cite{pierro2024mambaptq} and Quamba \cite{chiang2024quamba}. The former highlights potential challenges in quantizing Mamba, and the latter proposes effective methods for handling the quantization of SSM inputs and outputs. However, these works have not focused on optimizing vision tasks, which makes it difficult to leverage the specific characteristics of vision tasks. Additionally, they have not explored in detail how to manage the computations within the SSM, which overlooks the challenges associated with the quantization of intermediate values generated during recurrent inference.

In this paper, we propose a novel PTQ method for ViM. To our best knowledge, this is the first study to perform PTQ on ViM, including a detailed quantization of the SSM layers. Our main contributions are as follows:
\begin{itemize}  
    \item \textit{A similarity-based k-scaled token-wise quantization.} We propose a method for the quantization of linear layers that is specifically tailored to the characteristics of image processing. This approach effectively addresses the challenges caused by large variations between tokens.
    \item \textit{A reparameterization method for SSM.} We develop a reparameterization technique to "smoothing" the hidden state and make it easier to quantize. This method improves the accuracy of computations and reduces the impact of error propagation during quantization of SSM .
    \item \textit{Determining the reparameterization factor.} We propose a method for finding the suitable reparameterization factor based on tensor distribution. This approach reduces computational overhead by allowing part of the reparameterization computations to merge with preceding linear layers, while the selected factor ensures better accuracy.
\end{itemize}

\section{Related Works and Background} \label{Sec::Background}

\subsection{Preliminaries of SSM and ViM} \label{Sec::2.1}
Recently, the Mamba model \cite{gu2023mamba}\cite{dao2024transformers} has been proposed. Distinct from the widely used Transformer, Mamba utilizes selectivity in its SSM to reduce time complexity while achieving the same modeling power as the Transformer. Mamba model centers around the SSM architecture. This system maps a one-dimensional sequence $x_t$ to $y_t$ through a hidden state $h_t$. The system incorporates five parameters: $\mathbf{A}$, $\mathbf{B}$, $\mathbf{C}$, $\mathbf{D}$, and a discretizing parameter $\Delta$. The discretization operates as follows:
\begin{equation}
    \begin{split}
        \Bar{\mathbf{A}} & = \exp(\Delta \mathbf{A}), \\
        \Bar{\mathbf{B}} & = {\left(\Delta \mathbf{A}\right)}^{-1}\left(\exp\left(\Delta \mathbf{A}\right) - \mathbf{I}\right) \cdot \Delta \mathbf{B}.
    \end{split}
\end{equation}
The Mamba model introduces selectivity into the SSM system by making $\Delta$, $\mathbf{B}$, and $\mathbf{C}$ input-dependent. This modification enhances the model's ability to focus on relevant information and discard irrelevant ones. The resulting recurrent process can be expressed as:
\begin{equation}
    \begin{split}
        h_t & = \Bar{\mathbf{A}}h_{t - 1} + \Bar{\mathbf{B}}x_t, \\
        y_t & = \mathbf{C}h_t + \mathbf{D}x_t.
    \end{split}
\end{equation}

\begin{figure}[t]
    \centerline{\includegraphics[width=\linewidth]{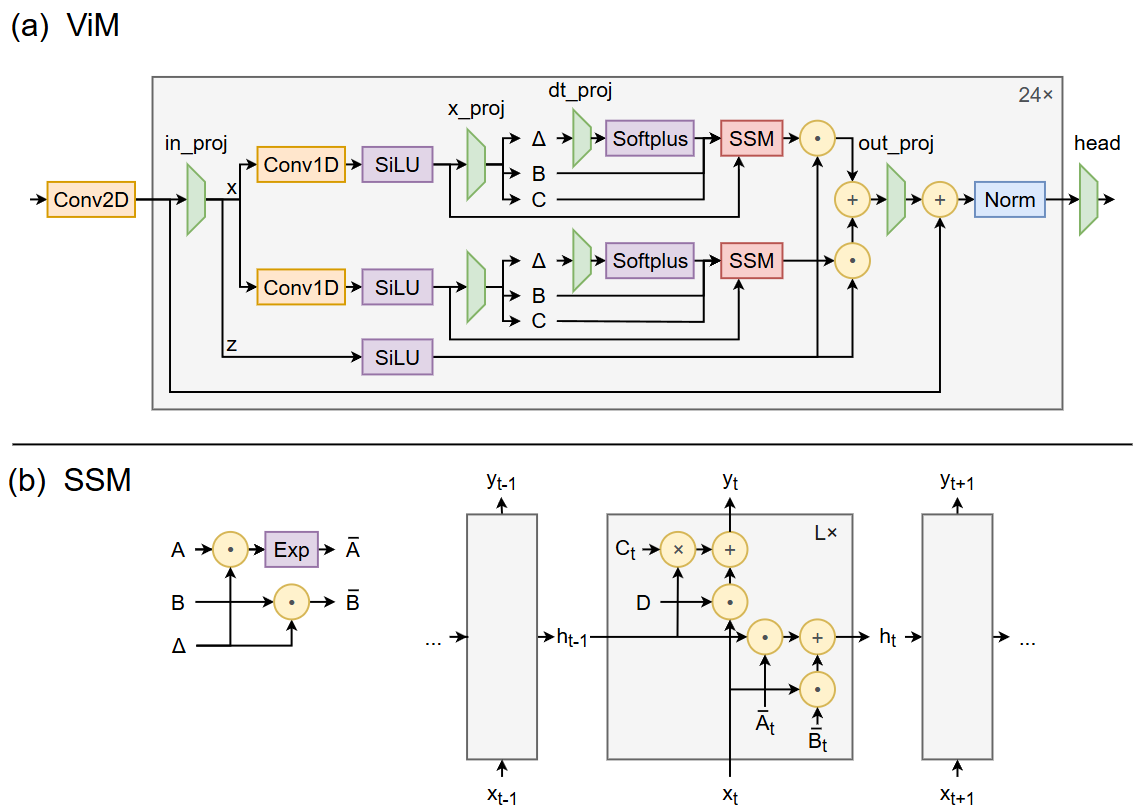}}
    \caption{(a) The architecture of ViM, with different types of layers represented in different colors. (b) Illustration of the operations within the SSM block.}
    \label{fig::ViM_architecture}
\end{figure}

As a versatile model backbone, numerous subsequent works have rapidly applied Mamba to vision tasks. Vision Mamba \cite{2024_ViM} employs bidirectional SSM architecture as shown in \cref{fig::ViM_architecture}. Aiming to enhance the handling of spatial dependencies within images. VMamba \cite{liu2024vmamba} proposed 2D Selective Scan with the same intention. EfficientVMamba \cite{pei2024efficientvmamba} introduce an atrous-based selective scan with efficient skip sampling to ensure global information while minimizing overhead. MambaVision \cite{hatamizadeh2024mambavision} further integrates Mamba blocks with attention blocks to achieve a balance between performance and accuracy, ensuring that the model can incorporate both sequential and spatial information.

\subsection{Preliminaries of Post-Training Quantization} \label{Sec::2.2}
PTQ is a technique used to reduce the size and computational complexity of a pre-trained neural network by converting its weights and activations to lower bitwidth. It uses a portion of the training data as calibration data to collect the outputs and intermediate states produced by the model when processing these inputs. This information is then used to quantize the weights and activations of the model. Its advantages, including the lack of a need for re-training and minimal data requirements, make it a widely adopted method for reducing model size and enabling efficient inference.

\textbf{Symmetric MinMax quantization} \cite{jacob2018quantization} is the most basic quantization method, and its quantizer can be defined as follows:
\begin{equation}
    Q(x) = clamp\left(\left\lfloor\frac{(2^{b - 1} - 1)x}{\max(|x|)}\right\rceil, -2^{b - 1}, 2^{b - 1} - 1 \right).
\end{equation}
Here, $clamp(x, l, h)$ denotes the operation of clamping $x$ within the range $[l, h]$. This method uses the maximum absolute value of the data to determine the quantization range, which ensuring that all input values fall within the range representable by the quantized format. In this paper, symmetric MinMax quantization is employed as our preliminary quantization method. For layers that are challenging to quantize directly using symmetric MinMax quantization, we propose specific methods in \cref{Sec::3} and \cref{Sec::4}.

\subsection{Related Works} \label{Sec::2.3}

\textbf{Related PTQ for Transformer} In the past, numerous PTQ methods have been proposed for Transformers. \cite{wu2020integer} summarized the fundamental methods for PTQ in neural networks and provided a comprehensive benchmark. These methods are commonly used as baseline approaches in many works. AdaRound \cite{nagel2020up}, OMSE \cite{choukroun2019low} and EasyQuant \cite{wu2020easyquant} proposed to adjust rounding and scale selecting process respectively to minimize quantization error. PTQ4ViT \cite{yuan2022ptq4vit}, FQ-ViT \cite{lin2021fq} and TSPTQ-ViT \cite{tai2023tsptq} address challenges such as extreme activation distributions and inter-channel variations by employing multi-scale quantization techniques. These techniques including twin uniform quantization in PTQ4ViT, Power-of-Two Factor in FQ-ViT and Two-Scaled Scaling Factor in TSPTQ-ViT. PTQ4ViT also utilizes a Hessian-guided metric to select optimal quantization scales; On the other hand, SmoothQuant \cite{xiao2023smoothquant}, AWQ \cite{lin2024awq} and RepQ-ViT \cite{li2023repq} focus on adjusting weights or parameters to simplify the quantization process. SmoothQuant applies per-channel smoothing factors to reduce activation dynamic ranges. AWQ introduces activation-aware weight-only quantization to further compress model size. RepQ-ViT employs scale reparameterization to transfer inter-channel variation in LayerNorm to linear layers; MPTQ-ViT \cite{tai2024mptq} further improves methods in SmoothQuant and TSPTQ-ViT, and proposes a loss function to efficiently determine the optimal mixed-precision bitwidth in a greedy manner.

\textbf{PTQ for Mamba-based Model.} Currently, there is limited research on PTQ for Mamba. Mamba-PTQ \cite{pierro2024mambaptq} conducted baseline quantization for Mamba and identified the main challenge of Mamba quantization as stemming from activation outliers. They proposed an outlier-aware quantization approach but were unable to fully address quantization errors. Quamba \cite{chiang2024quamba} tackled this issue by applying percentile quantization to the SSM inputs to eliminate extreme values and leveraging the Walsh–Hadamard transform on the outputs to manage large outliers. Their method outperformed other baseline quantization techniques in language tasks. However, these works have not focused on optimizing vision tasks, which limits their applicability to exploit the unique characteristics of vision tasks, such as structured patterns in data distributions. Additionally, these methods have not explored in detail how computations within the SSM can be managed during quantization. This oversight neglects the impact of intermediate values generated during recurrent inference, which cause significant error propagation in Mamba-based models.

\section{Similarity-based k-Scaled Quantization} \label{Sec::3}

% In this section, we focus on the quantization of convolutional layers and linear projection layers in ViM.

\subsection{Quantization Difficulty} \label{Sec::3.1}
The difficulties of quantization are the outlier and the large dynamic range. Both convolution and linear projection operations can be viewed as matrix multiplication:
\begin{equation}
    \mathbf{Y} = \mathbf{W} \cdot \mathbf{X}, 
\end{equation}
where we apply quantization to the model weights $\mathbf{W}$ and the layer outputs $\mathbf{Y}$. Under 8-bit 
symmetric MinMax quantization, while $\mathbf{W}$ does not have a significant impact; certain layers of $\mathbf{Y}$ exhibit considerable accuracy drops. The output distributions of these layers shows common characteristics: a large dynamic range and prominent outliers. A large dynamic range can make it challenging to efficiently represent values during quantization, as it requires a broader range of representable values, which leads to loss of precision. On the other hand, prominent outliers can dominate the quantization scale and leads to suboptimal quantization of the majority of the data.

The large dynamic range and outliers may stem from the characteristics of the ViM, which previous works were unable to address. Furthermore, in the past, these two issues were often treated as a single problem. However, in this work, we distinguish between their characteristics and tackle them separately. To handle the large dynamic range, we employ a similarity-based method; while for outliers, we apply a k-scaled quantization approach. Additionally, we integrate both the similarity-based and k-scaled methods to provide a more comprehensive solution.

\subsection{Similarity-based Quantization for Large Dynamic Range} \label{Sec::3.2}
\begin{figure}[t]
    \centerline{\includegraphics[width=\linewidth]{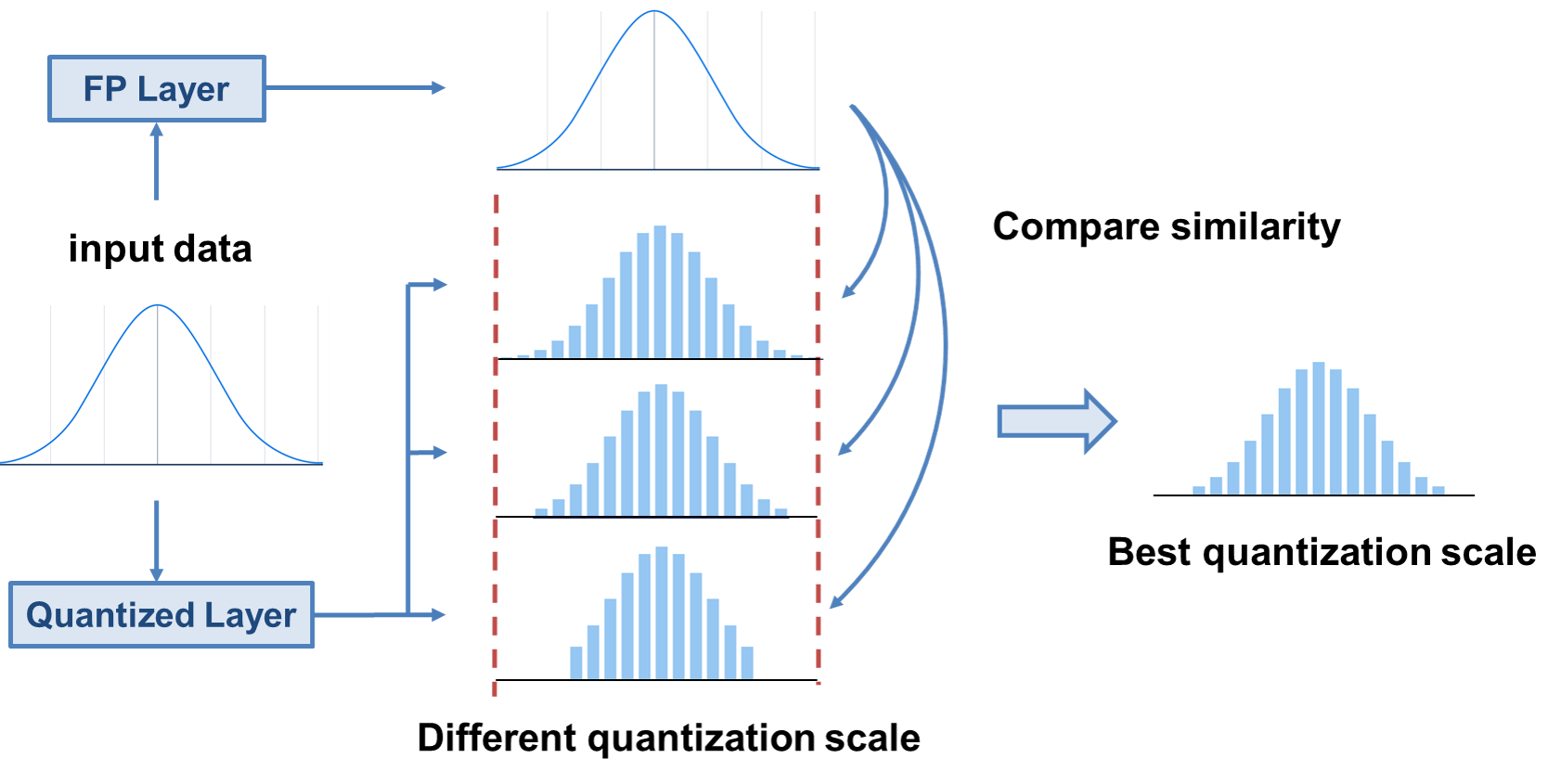}}
    \caption{Similarity-based quantization.}
    \label{fig::Similarity-based}
\end{figure}

The similarity-based method finds the scale to achieve a balance for values that has large dynamic range. This method ensures that the quantization process effectively captures the overall distribution without being overly influenced by extreme values. It has also been used in past research on PTQ, and it is a universal quantization method applicable across different layers. As shown in \cref{fig::Similarity-based}, the output values $y$ of each layer in the original floating-point (FP) model are first recorded during calibration. Then, quantization is performed using different quantization scales from a pre-defined search space, and the similarity between the quantized results $\hat{y}$ and the original values $y$ is computed. The goal is to find the scale with the highest similarity:
\begin{equation}
    \mathop{\arg\max}_{s \in \mathbf{S}}\left(\text{sim}\left(y, Q\left(y \middle| s\right) \cdot s\right)\right),\label{eq::similarity_based}
\end{equation}
where $\mathbf{S}$ is the search space of scale $s$, $\text{sim}(x, y)$ means similarity between $x$ and $y$, and $Q\left(y \middle| s\right)$ represent the quantized result of $y$ using scale $s$:
\begin{equation}
    Q\left(y \middle| s\right) = clamp\left(\left\lfloor\frac{x}{s}\right\rceil, -2^{b - 1}, 2^{b - 1} - 1 \right).
\end{equation}
The optimal scale found in \cref{eq::similarity_based} identifies a balanced scale for widely distributed values and is thus selected as the quantization scale during the inference process.

There are various similarity metrics to choose from, including cosine similarity, L1 norm, L2 norm, and the Hessian-guided metric \cite{yuan2022ptq4vit}. Our experimental results show that both cosine similarity and the Hessian-guided metric perform best. However, the Hessian-guided metric requires additional calibration overhead, so we will use cosine similarity as the default similarity-based quantization setting.

\subsection{k-Scaled Quantization for Outlier} \label{Sec::3.3}
The similarity-based quantization method effectively reduces quantization error in a general sense but remains inadequate in addressing extreme outliers within specific channels or tokens. Previous solutions to this issue include power-of-two factor from FQ-ViT \cite{lin2021fq}, and the two-scaled factor from TSPTQ-ViT \cite{tai2023tsptq}. The core concept behind these methods is to use distinct quantization scales for different channels so that both typical values and extreme outliers can be accurately captured.

Building on this concept, we propose two advanced methods: \textit{k-scaled channel-wise quantization} and \textit{k-scaled token-wise quantization}, designed to further enhance quantization accuracy by adapting to the unique characteristics of each channel or token. In k-scaled channel-wise quantization, when calibrating the activation $y$ whose channel containing outliers, we first partition all channels using the k-means clustering algorithm. Then, an optimal quantization scale is determined for each cluster:
\begin{equation}
    s_1 \geq s_2 \geq \cdots \geq s_k,
\end{equation}
This ensuring that the quantization process adapts to the specific characteristics of each channel. Then, for hardware-friendly computating, we redefine the quantization scale as:
\begin{equation}
    s_i^* = 
    \begin{cases}
        s_1, & \text{if $i = 1$} \\
        s_1 >> \left\lfloor\log\frac{s_i}{s_1}\right\rceil & \text{otherwise}
    \end{cases}
\end{equation}
where $s_1$ is the largest scale found in previous step, and $>>$ represents a bitwise right shift operation. This method allows each channel to use a distinct quantization scale, thereby improving quantization accuracy while ensuring efficient implementation in hardware.

\begin{figure}[t]
    \centerline{\includegraphics[width=\linewidth]{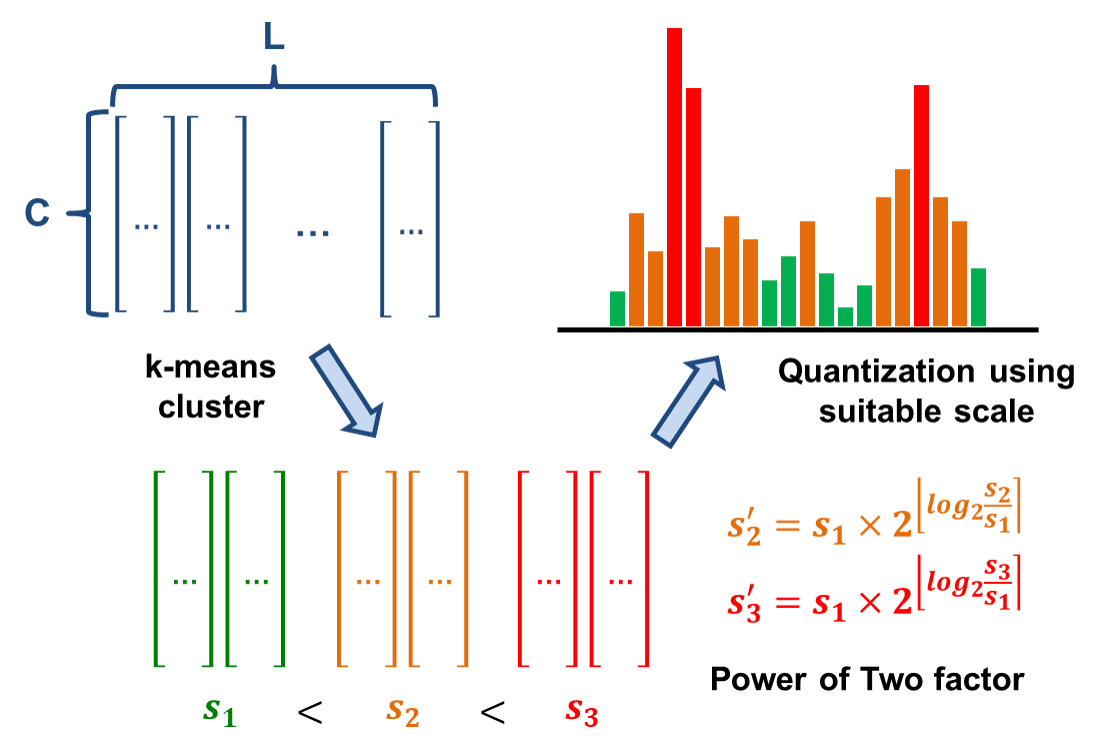}}
    \caption{k-scaled token-wise quantization.}
    \label{fig::k-Scaled_Token-wise}
\end{figure}

The k-scaled channel-wise quantization method effectively improves accuracy for the convolutional layers but shows limited effectiveness for some linear projection layers. Upon further investigation, we observe that the outliers in these layers are not strongly correlated with specific channels. Instead, they are associated with certain tokens that exhibit particularly extreme values. Additionally, we identify two critical facts:
\begin{itemize}
    \item For image processing tasks, the token length is determined by how the input image is segmented before being fed into the model. Unlike in natural language processing tasks, the token length in image processing tasks can be a fixed value.
    \item Across different images, the outliers in these linear layers consistently occur in the middle tokens. This indicates that the value distribution of calibration data over tokens is sufficiently representative, allowing the quantization scale to be determined based on their value distribution.
\end{itemize}
Based on these observations, we propose k-scaled token-wise quantization to address the quantization challenges of these linear layers. In \cref{fig::k-Scaled_Token-wise}, similar to the channel-wise approach, we apply k-means clustering to partition $y$ along the token dimension and redefine the quantization scale to achieve a hardware-friendly dequantization computation flow. This method effectively mitigates the issue of token outliers in linear projection layers.

\section{Quantization for SSM} \label{Sec::4}
The SSM is the key component of the Mamba architecture, providing Mamba-based models with the ability to selectively focus on or filter out inputs within a sequential state. Unlike traditional attention blocks, SSM achieves this with memory and computation requirements that scale linearly with the token length. In this section, we focus on the core SSM operations within ViM. Specifically, our goal is to quantize the hidden state $h$ during the recurrent process of the SSM.

\subsection{Quantization Difficulty} \label{Sec::4.1}
In SSM, the hidden state $h \in \mathbb{R}^{C\times D\times L}$ introduces an additional hidden dimension compared to the input and output of the SSM. This expansion leads to a more dispersed value distribution, with numerous extreme values occurring across different channels and hidden dimensions. These large values are the primary challenge for quantization, as they contribute significantly to the overall quantization error.

On the other hand, quantization in traditional deep neural networks often suffers from error propagation caused by quantization errors occurring in earlier stages. In SSM, this issue is further aggravated by the absence of non-linear activation functions, which typically serve to constrain the value distribution. This amplifies the quantization error stemming from the uneven distribution of the hidden state, as mentioned earlier. Consequently, developing an effective quantization method for SSMs is a critical challenge when quantizing ViM and other Mamba-based models.

\subsection{SSM Reparameterization} \label{Sec::4.2}
To address the above issue, we first observe the value distribution of the hidden state. While the differences in hidden state values across different layers of the SSM are quite large, they exhibit a certain form of ``regularity.'' Specifically, when we examine the values along the channel dimension, we find that two distinct channels have similar distributions, differing primarily in scale. This observation can be expressed mathematically in some sense as:
\begin{equation}
    h[i, :, :] \approx r \cdot h[j, :, :],
\end{equation}
where $i$ and $j$ refer to different channels, and $r$ is a scaling factor. Note that this equation is not strictly correct, but rather reflects the relationship identified through observation. Interestingly, this relationship can also be observed along the token and hidden dimension directions.

\begin{figure}[t]
    \centerline{\includegraphics[width=\linewidth]{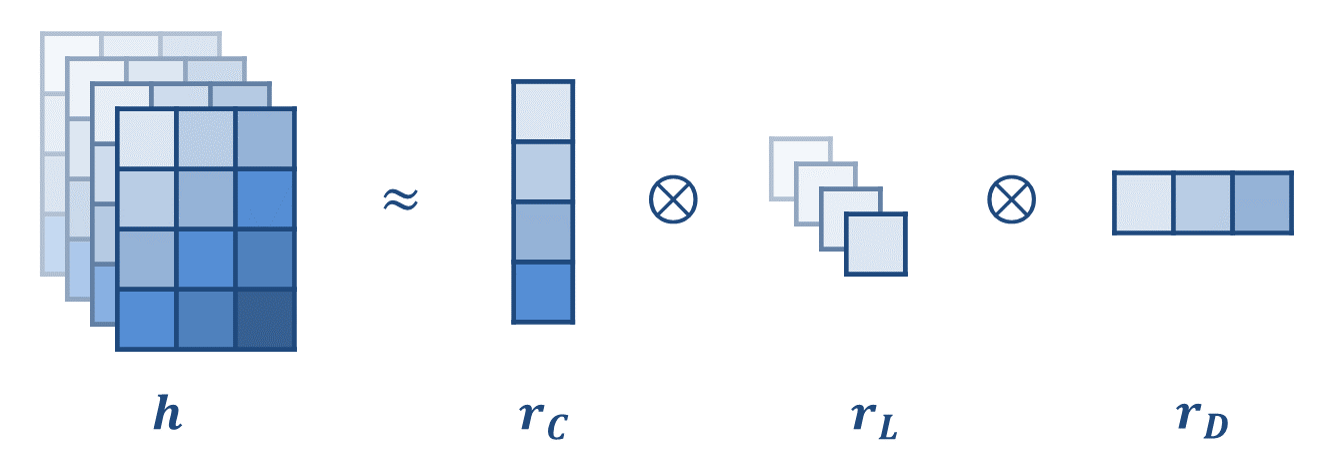}}
    \caption{Rank-1 approximation of $h_t$.}
    \label{fig::Rank1_approximation}
\end{figure}

Based on this crucial observation, we find that the hidden state $h$ can be effectively approximated by a rank-1 approximation using three vectors as shown in \cref{fig::Rank1_approximation}: $r_C \in \mathbb{R}^C$, $r_L \in \mathbb{R}^L$, $r_D \in \mathbb{R}^D$, such that:
\begin{equation}
    h \approx r_C \otimes r_L \otimes r_D,
\end{equation}
where $\otimes$ denotes the tensor product.

The reason for approximating $h$ is to smooth the value distribution of $h$. The uneven distribution of $h$ creates significant difficulties in quantization. Hence, by smoothing $h$ before quantization, the range of its values becomes narrower, enabling the use of more precise quantization bins and thereby improving the quantization accuracy. Therefore, our objective is to divide $h$ by its approximation to obtain:
\begin{equation}
    h^* = \frac{h}{r_C \otimes r_L \otimes r_D}.
\end{equation}
Alternatively, for a specific token $t$, we can express it as:
\begin{equation}
    h_t^* = \frac{h_t}{r_C \otimes {r_L}_t \otimes r_D}.
\end{equation}
By performing this operation, we obtain $h^*$, which exhibits a smoother value distribution, making it easier to quantize to $\hat{h}^*$ afterward with higher accuracy.

To avoid the additional computational overhead that could outweigh the optimization benefits of quantization, we propose a method of reparameterization. By ``reparameterizing'' the parameters $\Bar{\mathbf{A}}$, $\Bar{\mathbf{B}}$, and $\mathbf{C}$ with the reparameterization factors $r_C$, $r_L$, and $r_D$, the need of additional multiplication and division at each recurrent step can be eliminated. We begin by analyzing the operations performed in the SSM:
\begin{equation}
    \begin{split}
        h_t & = \Bar{\mathbf{A}}_t \odot h_{t - 1} + \Bar{\mathbf{B}}_t \odot x_t, \\
        y_t & = \mathbf{C}_t \times h_t + \mathbf{D} \odot x_t. \\
    \end{split}
\end{equation}
Here, $x, y \in \mathbb{R}^{C \times L}$, $\Bar{\mathbf{A}}, \Bar{\mathbf{B}}, h \in \mathbb{R}^{C \times L \times D}$, and $\mathbf{C} \in \mathbb{R}^{L \times D}$. Additionally, since $\mathbf{D}$ does not change with respect to $t$, the $\mathbf{D} \odot x_t$ term can be added before the output of the SSM. Therefore, we only need to consider the remaining terms for further analysis and optimization:
\begin{equation}
    y_t = \mathbf{C}_t \times h_t.
\end{equation}
As mentioned previously, at the intermediate step $i$, we aim to smooth $h_i$ using $r_C \otimes {r_L}_i \otimes r_D$. Therefore, at $t = 1$, since $h_0 = 0$, we can reparameterize $\Bar{\mathbf{B}}$ as follows:
\begin{equation}
    \begin{split}
        h_1^* & = \frac{h_1}{r_C \otimes {r_L}_1 \otimes r_D} \\
        & = \Bar{\mathbf{A}}_1 \odot 0 + \frac{\Bar{\mathbf{B}}_1}{r_C \otimes {r_L}_1 \otimes r_D} \odot x_1.
    \end{split}
\end{equation}
At subsequent steps, instead of multiplying $h_t^*$ back to recover $h_t$ and introducing significant computational overhead, we directly use $h_t^*$ in the following calculations. However, this introduces numerical errors. To address this, we reparameterize $\Bar{\mathbf{A}}$ as follows:
\begin{equation}
    \begin{split}
        h_t^* & = \frac{h_t}{r_C \otimes {r_L}_t \otimes r_D} \\
        & = \left(\Bar{\mathbf{A}}_t \odot \frac{{r_L}_{t - 1}}{{r_L}_t}\right) \odot h_{t - 1}^* + \frac{\Bar{\mathbf{B}}_t}{r_C \otimes {r_L}_t \otimes r_D} \odot x_t.
    \end{split} \label{eq::h_star}
\end{equation}
Similarly, when computing the output, we directly use $h_1^*$ to calculate the result and correct any errors through reparameterization of $\mathbf{C}$ as follows:
\begin{equation}
    \frac{y_t}{r_C} = \left(\mathbf{C} \odot ({r_L}_t \otimes r_D)\right) \times h_t^*.
\end{equation}
Summarizing the reparameterization process, the required transformations are as follows:
\begin{equation}
    \Bar{\mathbf{A}}^* = \Bar{\mathbf{A}} \odot \left[1, \frac{{r_L}_1}{{r_L}_2}, \cdots, \frac{{r_L}_{L - 1}}{{r_L}_L}\right], \label{eq::DeltaAReparam}
\end{equation}
\begin{equation}
    \Bar{\mathbf{B}}^* = \Bar{\mathbf{B}} \odot \frac{1}{r_C \otimes r_L \otimes r_D},\label{eq::DeltaBReparam}
\end{equation}
\begin{equation}
    \mathbf{C}^* = \mathbf{C} \odot (r_L \otimes r_D).\label{eq::CReparam}
\end{equation}
Finally, after obtaining $y_t / r_C$ for each step, we perform a multiplication to recover the original $y$:
\begin{equation}
    y = \left[\frac{y_1}{r_C}, \frac{y_2}{r_C}, \cdots, \frac{y_L}{r_C}\right] \odot r_C.\label{eq::yReparam}
\end{equation}

\begin{figure*}[tb]
    \centerline{\includegraphics[width=\linewidth]{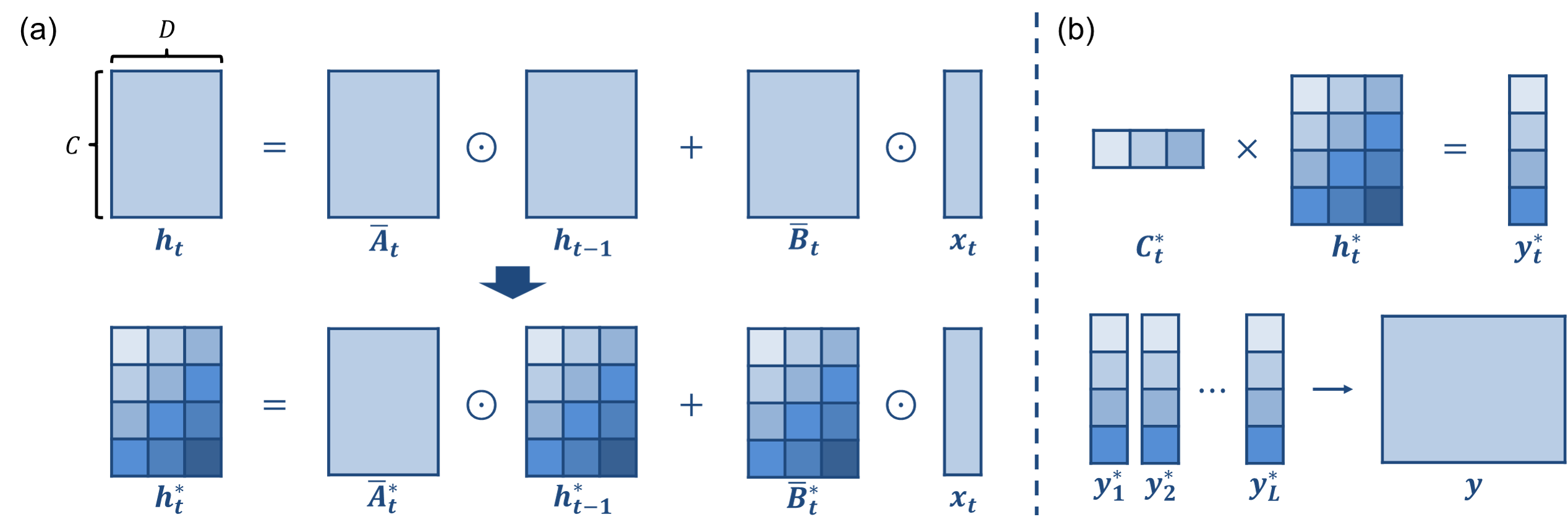}}
    \caption{Reparameterization diagram. (a) Illustration of the difference between the calculation of \( h_t^* \) using \cref{eq::h_star} and the original SSM calculation of \( h_t \), highlighting how the reparameterized process modifies intermediate computations while preserving equivalence. (b) demonstration of how \( h_t^* \) is used at each step to compute \( y_t^* \) and how \( y_t^* \) is eventually transformed back into the original SSM output \( y \), ensuring consistency with the original formulation.}
    \label{fig::reparameterization}
\end{figure*}
\cref{eq::DeltaAReparam,eq::DeltaBReparam,eq::CReparam,eq::yReparam} also explains the reason behind employing rank-1 approximation to approximate $h$ as introduced earlier. By representing $h$ as a tensor product of three vectors $r_C$, $r_L$, and $r_D$, we are able to capture its essential structure while significantly reducing the complexity of its value distribution. This approach enabling more precise quantization with reduced computational overhead, ensuring efficient implementation and accuracy in the final output recovery.

\subsection{Factorization for Computing Reduction} \label{Sec::4.3}
As discussed in \cref{Sec::4.2}, the parameters $\Bar{\mathbf{A}}$, $\Bar{\mathbf{B}}$, and $\mathbf{C}$ need to be reparameterized as shown in \cref{eq::DeltaAReparam,eq::DeltaBReparam,eq::CReparam}. Here, we demonstrate that the computational overhead during inference can be further reduced by fusing the multiplication of $\Bar{\mathbf{B}}$ and $\mathbf{C}$ into the linear layer before entering the SSM. 
The calculation process for $\Bar{\mathbf{B}}$ and $\mathbf{C}$ is as follows:
\begin{equation}
    \Bar{\mathbf{B}} = \Delta \odot B = Softplus(\mathbf{W_\Delta} \cdot x) \odot (\mathbf{W_B} \cdot x),
\end{equation}
\begin{equation}
    \mathbf{C} = \mathbf{W_C} \cdot x.
\end{equation}
Note that both $\mathbf{B}$ and $\mathbf{C}$ are derived from linear operation. This means that the multiplication in \cref{eq::DeltaBReparam,eq::CReparam} can be moved forward and computed together with the projection layer. We can modify the weights $\mathbf{W_B}$ and $\mathbf{W_C}$ as follows:
\begin{equation}
    \mathbf{W_B}^* = \mathbf{W_B} \odot \frac{1}{r_D},\  \mathbf{W_C}^* = \mathbf{W_C} \odot r_D.
\end{equation}
As a result, \cref{eq::DeltaBReparam,eq::CReparam} can be rewritten as:
\begin{equation}
    \Bar{\mathbf{B}}^* = \Bar{\mathbf{B}} \odot \frac{1}{r_C \otimes r_L},\  \mathbf{C}^* = \mathbf{C} \odot r_L.
\end{equation}

Reparameterization factors $r_C$, $r_L$, and $r_D$ are required for the rank-1 approximation of $h$. Since our primary goal is to transform $h$ into a smoother version $h^*$, the scaling factors $r_C$, $r_L$, and $r_D$ can be arbitrarily scaled by a constant factor without impacting the effectiveness of reparameterization. Therefore, these factors are only required to represent the variation trends of $h$ across each dimension.

A reasonable initial choice for the reparameterization factors is to compute the representative of $h$ along each corresponding dimension as , where "$\mathrm{Rep}$" is the adjustable representative function (e.g., mean) for $h$:
\begin{equation}
    {r_C}_0 = \underset{0 \leq l < L, 0 \leq d < D}{\mathrm{Rep}}h_{c, l, d},
\end{equation}
\begin{equation}
    {r_L}_0 = \underset{0 \leq c < C, 0 \leq d < D}{\mathrm{Rep}}h_{c, l, d},
\end{equation}
\begin{equation}
    {r_D}_0 = \underset{0 \leq c < C, 0 \leq l < L}{\mathrm{Rep}}h_{c, l, d}.
\end{equation}
Since the impact of each dimension on $h$ varies, the adjustments to the reparameterization factors should reflect the distribution of $h$ along each dimension. A dimension with higher variability should exert more influence. Consequently, we adjust $r_C$, $r_L$, and $r_D$ by weighting them with their respective dispersion, where "$\mathrm{Disp}$" is the adjustable dispersion function (e.g., standard deviation):
\begin{equation}
    r_C = pow\left({r_C}_0, \frac{\mathrm{Disp}\left({r_C}_0\right)}{\mathrm{Disp_{sum}}}\right),
\end{equation}
\begin{equation}
    r_L = pow\left({r_L}_0, \frac{\mathrm{Disp}\left({r_L}_0\right)}{\mathrm{Disp_{sum}}}\right),
\end{equation}
\begin{equation}
    r_D = pow\left({r_D}_0, \frac{\mathrm{Disp}\left({r_D}_0\right)}{\mathrm{Disp_{sum}}}\right).
\end{equation}
Here, $pow(x, a)$ denotes $x$ raised to the power of $a$, and $\mathrm{Disp_{sum}} = \mathrm{Disp}\left({r_C}_0\right) + \mathrm{Disp}\left({r_L}_0\right) + \mathrm{Disp}\left({r_D}_0\right)$

\section{Experimental Result} \label{Sec::Exp}
% In this section, we present the experimental results of our proposed quantization methods. 

\subsection{Experimental Setups} \label{Sec::5.1}
\textbf{Models and datasets} We conduct experiments on an image classification task, focusing on the quantization of ViM. We use pretrained models provided by the original authors, including ViM-T and ViM-S, both of which are pretrained on the ImageNet-1k dataset \cite{deng2009imagenet}. ImageNet-1k contains 1.28M training samples and 50K validation samples across 1,000 categories. For calibration, we randomly sample 256 training samples from the training set and use the validation set to evaluate performance.

\textbf{Implementation details} For the quantization setup, we quantize all linear and convolutional layers, along with the SSM block in the model. We use symmetric MinMax quantization as our baseline quantization method. For the linear and convolutional layers, we apply Similarity-based and k-scaled quantization as described in \cref{Sec::3} for the \texttt{conv1d} and \texttt{out\_proj} layers, and use symmetric layer-wise MinMax quantization for other layers for simplicity. The hyperparameter $k$ in k-scaled quantization is set to $4$. For the SSM block, we apply the reparameterization technique proposed in \cref{Sec::4}.

\subsection{Ablation Experiment} \label{Sec::5.2}
\begin{figure}[tb]
    \centerline{\includegraphics[width=\linewidth]{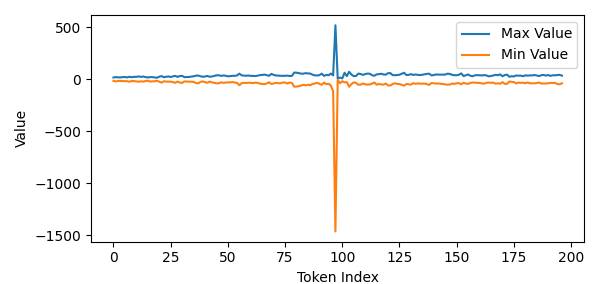}}
    \caption{Large outlier in output of the output projection layer.}
    \label{fig::outlier}
\end{figure}

\textbf{Similarity-based \& k-scaled quantization.} We conduct ablation experiments on similarity-based and k-scaled quantization methods, where only the linear and convolutional layers are quantized. We found that the \texttt{conv1d} and \texttt{out\_proj} layers in ViM contribute significantly to quantization errors, resulting in accuracy drops of $2.9\%$ and $10.2\%$, respectively. Our observations reveal that the outputs of these layers contain large dynamic range and substantial outliers, as illustrated in \cref{fig::outlier}.
We proposed similarity-based and k-scaled method in \cref{Sec::3.2} and \cref{Sec::3.3} to addressed this issue. The results are shown in \cref{tab::similarity_k-scaled_result}. It can be observed that both similarity-based and k-scaled quantization methods demonstrate decent accuracy recovery capabilities. When combined, they further improve accuracy, reducing the gap with the FP model to just $0.6\%$, highlighting the effectiveness of these two methods.

\begin{table}[tb]
    \caption{Similarity-Based \& k-Scaled Quantization Result.}
    \begin{center}
        \begin{tabular}{llr|r}
            \toprule
            Model & Method & W/A & ImageNet-1k \\
            \midrule
            \multirow{5}{*}{ViM-T} & FP & $32/32$ & $76.1$ \\
            \cmidrule{2-4}
            & Baseline & $8/8$ & $62.2$ \\
            \cmidrule{2-4}
            & Similarity-Based & $8/8$ & $74.1$ \\
            & k-Scaled & $8/8$ & $75.3$ \\
            & Both & $8/8$ & $75.5$ \\
            \bottomrule
        \end{tabular}
        \label{tab::similarity_k-scaled_result}
    \end{center}
\end{table}

\begin{figure}[tb]
    \centerline{\includegraphics[width=\linewidth]{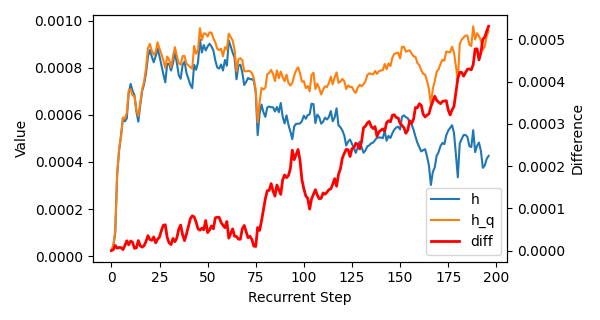}}
    \caption{Error propagation in SSM recurrent process.}
    \label{fig::error_propagation}
\end{figure}

\textbf{SSM Reparameterization.} We also conducted ablation experiments on the SSM reparameterization method. The quantization challenge in SSM arises from severe error propagation. As shown in \cref{fig::error_propagation}, the difference between the original $h$ and the quantized $\hat{h}$ increases significantly as the recurrent process progresses.
To address this issue, we proposed a reparameterization method in \cref{Sec::4.2} aimed at mitigating the error propagation effect. For the choices of reparameter factor, we used the max, mean, and median values as representatives, and constant, standard deviation, and variance as dispersion measures. The results obtained under different reparameterization factor are shown in \cref{tab::reparameterization_factor}. Based on the results, we select the mean value as the representative and the standard deviation as the dispersion measure in the following experiments.

\begin{table}[tb]
    \centering
    \caption{Summary of Metrics Across Different Reps and Disps.}
    \begin{tabular}{l|rrrr}
        \toprule
        Rep / Disp & Same & Std & Var & Range \\ 
        \midrule
        Max & $72.3$ & $69.1$ & $67.0$ & $73.1$ \\ 
        Mean & $72.8$ & $\mathbf{75.3}$ & $75.2$ & $75.0$ \\ 
        Median & $72.1$ & $75.2$ & $74.9$ & $74.9$ \\ 
        Percentile & $72.4$ & $69.0$ & $67.1$ & $73.2$ \\ 
        Quantile & $72.7$ & $75.2$ & $75.3$ & $75.3$ \\ 
        \bottomrule
    \end{tabular}
    \label{tab::reparameterization_factor}
\end{table}

\subsection{Experimental Results} \label{Sec::5.3}
The overall quantization results of image classification on the ImageNet-1k dataset for the quantized ViM model are presented in \cref{tab::overall_result}. The baseline method which employs symmetric MinMax quantization struggles to maintain model accuracy, particularly for the ViM-T model, where the accuracy drops dramatically to $12.2\%$. Similarly, for the ViM-S model, the baseline method results in an accuracy of $51.6\%$, which is still significantly lower compared to the original FP model. In contrast, our proposed quantization method significantly narrowing the accuracy gap between the quantized and FP models. For the ViM-T model, our method achieves an accuracy of $74.9\%$, only a $1.2\%$ drop compared to the FP model's accuracy. Similarly, for the ViM-S model, our method reaches an accuracy of $79.7\%$, only a $0.8\%$ decrease from the FP model.

\begin{table}[tb]
    \caption{Quantization Results of Image Classification on ImageNet-1k Dataset.}
    \begin{center}
        \begin{tabular}{llr|r}
            \toprule
            Model & Method & W/A & ImageNet-1k \\
            \midrule
            \multirow{3}{*}{ViM-T} & FP & $32/32$ & $76.1$ \\
            \cmidrule{2-4}
            & Baseline & $8/8$ & $12.2$ \\
            \cmidrule{2-4}
            & Ours & $8/8$ & $74.9$ \\
            \midrule
            \multirow{3}{*}{ViM-S} & FP & $32/32$ & $80.5$ \\
            \cmidrule{2-4}
            & Baseline & $8/8$ & $51.6$ \\
            \cmidrule{2-4}
            & Ours & $8/8$ & $79.7$ \\
            \bottomrule
        \end{tabular}
        \label{tab::overall_result}
    \end{center}
\end{table}

\section{Conclusion} \label{Sec::Conclusion}
In this paper, we address the challenges of quantizing ViM and highlight the difficulties in quantizing different layers. To improve accuracy, we apply the Similarity-based Method and propose a new k-scaled token-wise quantization method, both of which work for linear and convolutional layers. For SSM quantization, we introduce a reparameterization method that effectively handles the outlier issue in hidden state quantization. We also propose an appropriate reparameterization factor and combine the extra computations into the linear layer to reduce computational overhead. Experimental results show that our approach enables effective quantization of ViM while maintaining performance.

%% The file named.bst is a bibliography style file for BibTeX 0.99c
\bibliographystyle{IEEEbib}
\bibliography{refs}

\end{document}